\documentclass[aps,prl,amsmath,twocolumn,amssymb,superscriptaddress]{revtex4}
\pdfoutput=1
\usepackage{graphicx}
\usepackage{epstopdf}
\usepackage{color}
\usepackage{cases,array}

\begin{document}
\title{Phase Locked Loop of the Spin-Torque Nanooscillator}

\author{A.A. Mitrofanov}
\affiliation{National Research University "Moscow Power Engineering Institute", 112250 Moscow, Russia}

\author{A.R. Safin}
\affiliation{National Research University "Moscow Power Engineering Institute", 112250 Moscow, Russia}

\author{N.N. Udalov}
\affiliation{National Research University "Moscow Power Engineering Institute", 112250 Moscow, Russia}

\begin{abstract}
We have studied the dynamics of a spin-torque nanooscillator's (STNO) phase locked loop (PLL) generating microwave oscillations in a broad range of frequencies under the effect of direct current and external magnetic field. Bifurcations in the system caused by a change in the frequency detuning of synchronized oscillations are considered. Bands of phase locking and synchronism are determined. The existence of a phase locking band in the filterless PLL of STNOs basically distinguishes these oscillators from other types of microwave generators.
\end{abstract}

\maketitle

In spintronics, which is a new area in electronics, a key role is played by the intrinsic magnetic moment of electron spin, rather than by the electron charge. An important step in the development of spintronics was marked by the Slonczewski [1] and Berger [2], which showed the possibility of generating microwave oscillations by passing electric current due to spin transfer from one layer to another in a sample comprising alternating magnetic and nonmagnetic layers. This field of research has become very popular, and these microwave generators have received the special name of spin-torque nanooscillars (STNO).

Besides the obvious advantages related to their supersmall dimensions and broad range of frequency tuning under the action of magnetic field and electric current, modern STNOs still have some disadvantages, including low output power and large width of spectral lines. One possible method of decreasing the spectral line width of STNO oscillations is based on the mechanism of phase synchronization. Experiments [3] confirmed that, using a phase locked loop (PLL) system, it is possible to reduce the STNO linewidth. The first theoretical work devoted to the phase locking of STNOs was that by Mishagin and Shalfeev [4], in which various regimes of a PLL with an integrating filter in the control chain were considered; it was shown that the bandwidth of synchronous regimes in this case exceeds that for an STNO under direct action of an external harmonic signal; and the existence of a phase locking band in a filterless PLL of STNO was predicted.

In the present study, bifurcations in a filterless PLL of STNO caused by a change in the frequency detuning between oscillations of the STNO and a reference oscillator are considered and bands of phase locking and synchronism are determined.

The principle of PLL operation in STNOs has been described in detail in [4]. Let us consider processes in this system using the corresponding mathematical models. In contrast to the approach used in [4], which was based upon direct integration of the Landau-Lifshitz-Hilbert equation with an additional Slonczewski-Berger term [1, 2], it is more convenient to study STNOs based on the Slavin-Tiberkevich model, the applicability of which is discussed in detail in [5]. In order to obtain a model of PLL in STNO, the Slavin-Tiberkevich model is modified so as to take into account the influence of the error current signal related to the phase difference between the STNO and a reference oscillator:

\begin{equation}
\frac{\mathrm{d}c}{\mathrm{d}t} + j\omega(\vert c \vert ^2)c + \Gamma_G (1+Q\vert c \vert ^2)c - \sigma I (1-\vert c \vert ^2)c = 0. \label{eq:commonAdler}
\end{equation}
Here, $Q$ is the coefficient of positive nonlinear damping, $I=I_0+\Delta I=I_0(1+\epsilon K(p)F(\Delta \phi))$ is the STNO current, $K(p)$ is the operator filter transmission coefficient in the control chain, $F(\Delta \phi)$ is the normalized discrimination characteristic of the phase detector, $p=d/dt$ is the operator of differentiation, $\epsilon = \Delta I_{max}/I_0$, $\sigma$ is the spin transfer coefficient, $\omega(\vert c \vert ^2)$ is the STNO frequency defined as

\begin{equation}
\omega (\vert c \vert ^2) = \omega_0 + N \vert c \vert ^2,
\end{equation}
$\omega_0$ is the STNO ferromagnetic resonance frequency dependent on the magnetic field in the structure, $\vert c \vert ^2$ is the spin wave power, $N$ is the coefficient of the dependence of frequency on the square amplitude of the spin wave (for more detail, see [1]), $\Gamma_G=\alpha_G \omega_0$ is the coefficient of spin-wave positive losses in the free layer, and $\alpha_G$ is the linear damping coefficient.

\begin{figure*}[ht!]
\centerline{\includegraphics[width=120mm]{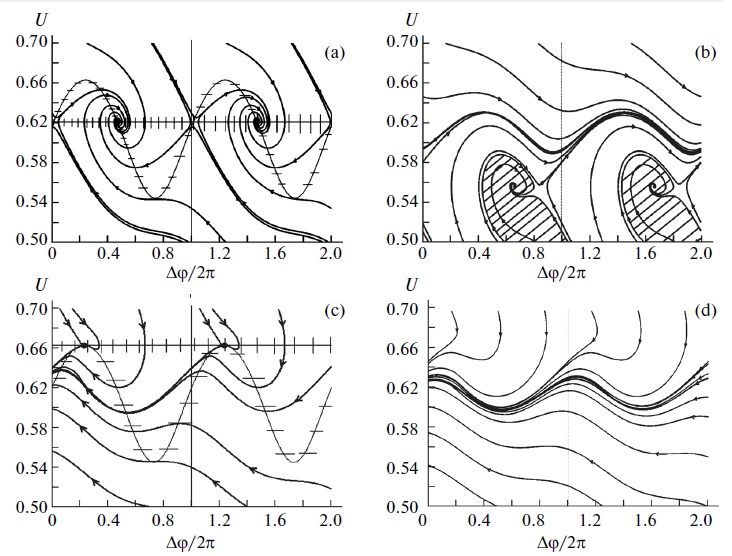}}
\caption{\footnotesize{Phase portraits of the model system with $N = 10.48 GHz$, $ \omega_0/2\pi= 12.41 GHz$, $Q = 0.66$, $\sigma = 61.5 GHz/A$, $\epsilon = 0.3$, $ \alpha_G = 0.01$, and $g=2$: (a) synchronous regime ($\Delta\omega/2\pi = –4.04 GHz$), (b) quasi-synchronous regime ($\Delta\omega/2\pi = –4.44 GHz$), (c) saddle-node bifurcation ($\Delta\omega/2\pi = –4.61 GHz$), and (d) regime of beats ($\Delta\omega/2\pi = –4.84 GHz$).}}
\label{fig:fig1}
\end{figure*}

Based on model (1), a filterless PLL with $F(\Delta \phi)=\sin(\Delta \phi)$ can be described by a system of equations with respect to slowly varying oscillation amplitude $U(t)$ and phase $\phi(t)$ of the STNO spin-wave with allowance for the influence of the phase synchronization chain. This model system can be written as follows:

\begin{subequations}
\begin{gather}\label{Trunc}
\frac{dU}{d\omega_0 t}=U\cdot \Gamma_G [(\zeta-1+\epsilon \cdot \sin(\Delta\phi)) - \\
- (\zeta+Q+\epsilon\sin(\Delta\phi))U^2]; \nonumber \\
\frac{d\Delta\phi}{d\omega_0t}=\Delta\omega+N\cdot U^2.
\end{gather}
\end{subequations}
where $\Delta\phi=\phi-\phi_0$,$\Delta\omega=\omega-\omega_0$  and $\phi_0$ and $\omega_0$ are the phase and frequency, respectively, of the reference oscillator. Thus, $\Delta\omega$ determines the difference of ferromagnetic resonance frequencies of the STNO free layer and the reference oscillator. If these frequencies are equal, the right-hand parts of Eqs.(3) vanish and, hence, $\Delta\omega=-N(U^0)^2$, where $U^0$ is the stationary amplitude of STNO oscillations.

Let us analyze system (3) by the phase space method. The equation of tangents to phase trajectories of the system can be written as follows:

\begin{gather}\label{Trunc}
\frac{dU}{d\Delta\phi}=\frac{U\cdot\Gamma_G[\zeta-1+\epsilon\sin(\Delta\phi)]}{\Delta\omega+N\cdot U^2}+ \\
+\frac{U^3\cdot\Gamma_G(\zeta+Q+\epsilon\sin(\Delta\phi))}{\Delta\omega+N\cdot U^2}.\nonumber
\end{gather}

A specific feature of this system is that the ordinate represents the amplitude of oscillations rather than the first derivative of the phase difference (which is a usual practice in the PLL theory for macroscopic autooscillators) [6]. However, the phase portraits are still similar to those for the classical system of phase synchronization. Stationary synchronous regimes correspond to the points of intersection of the isoclines of horizontal and vertical tangents. The isocline of the vertical tangent corresponds to stationary amplitude $U^0$, while the isocline of the horizontal tangent corresponds to a $2\pi$ periodic function and abscissa axis. The stationary values of nonzero amplitude and phase difference depend on the initial frequency difference.

Figure 1a presents a phase portrait corresponding to a synchronous regime, which shows two singular points (stable focus and saddle). An increase in the frequency detuning leads to bifurcation of confluence of the input and output separatrices of the saddle with the formation of a semistable limit cycle of the second kind. This value of detuning corresponds to a frequency difference equal to the locking bandwidth. With a further increase in the frequency detuning, the semistable limit cycle transforms into a stable limit cycle (Fig. 1b), while the input separatrix forms a loop (crosshatched region) bounding the domain of initial conditions for which the system comes into a stable sate. Finally, if the frequency detuning grows further and becomes equal to the bandwidth of synchronous regimes, the system exhibits a saddle–node bifurcation that corresponds to confluence of the stable and unstable singular points (Fig. 1c). When the frequency detuning exceeds the bandwidth of synchronous regimes, the isoclines of horizontal and vertical tangents do not intersect and the system passes to an asynchronous regime in which phase shift grows and the amplitude oscillates at a frequency of beats (Fig. 1d).

Using model system (3), it is possible to obtain the following expressions for stationary values of the phase and amplitude of oscillations:

\begin{subequations}
\begin{gather}\label{Trunc}
(U^0)^2=\frac{\zeta-1+\epsilon\sin(\Delta\phi^0)}{\zeta+Q+\epsilon\sin(\Delta\phi^0)}; \\
\Delta\phi^0=\arcsin[\frac{(\zeta-1)-\gamma(\zeta+Q)}{\zeta(\gamma-1)}],
\end{gather}
\end{subequations}
where $\gamma=\Delta\omega/N$ is the generalized frequency detuning. Substituting expression (5) into the second equation of system (3) and taking into account that $d\Delta\phi^0=0$, we obtain the following relation:

\begin{equation}
\Delta\omega=-N\frac{\zeta-1+\epsilon\sin(\Delta\phi^0)}{\zeta+Q+\epsilon\sin(\Delta\phi^0)}.
\end{equation}

Using this equation, it is possible to determine the band of synchronism. It is a specific feature of this system that the phase-discriminator signal leads to a change in the amplitude and frequency of STNO oscillations. This is a manifestation of the nonisochronism of these oscillators, which results in the existence of a phase-locking band in the filterless PLL. The band of synchronism is asymmetric for different signs of the initial frequency detunings between the STNO and reference oscillator. Indeed, Eq. (6) yields the following expressions for positive and negative detunings:

\begin{subequations}
\begin{gather}\label{Trunc}
\Delta\omega_{s+} = N\frac{\zeta-1+\epsilon}{\zeta+Q+\epsilon}-N\frac{\zeta-1}{\zeta+Q}; \\
\Delta\omega_{s-} = N\frac{\zeta-1-\epsilon}{\zeta+Q-\epsilon}-N\frac{\zeta-1}{\zeta+Q}.
\end{gather}
\end{subequations}
The main contribution to the bandwidths of synchronism and phase locking is related to parameter $\epsilon$, which is the maximum value of the normalized error signal.

\begin{figure}[ht!]
\centerline{\includegraphics[width=60mm]{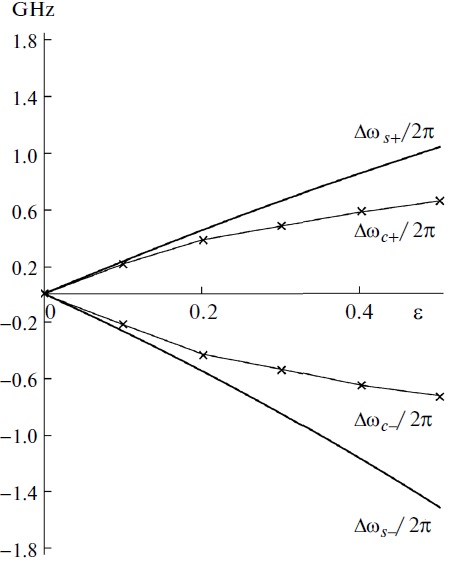}}
\caption{\footnotesize{Plots of synchronism bandwidth $\Delta\omega_s$ and phase-locking bandwidth $\Delta\omega_c$ vs. phase synchronization parameter $\epsilon$.}}
\label{fig:fig2}
\end{figure}

Figure 2 shows synchronism bandwidth $\Delta\omega_s$ in accordance with Eqs. (7a) and (7b) and phase-locking bandwidth $\Delta\omega_c$ in accordance with modeling results as functions of the phase-synchronization parameter $\epsilon$. The choice of maximum å is restricted to the linear region of the control characteristic of STNO, in which the amplitude of oscillations is sufficient for operation of the phase discriminator. In addition, the results of modeling show that an increase in å leads to increasing time required for attaining synchronism, which is an important factor in solving practical problems of constructing phase-synchronization systems for STNOs. Authors acknowledge financial support by Russian Foundation for basic research (Grant No. 10-02-01403) and Russian Federal Program "Scientific and scientific-pedagogical Staffs of Innovative Russia" for 2009-2013 (Grant No. 14.B37.21.1211, 14.132.21.1665).

\end{document}